\documentclass[11pt]{article}
\usepackage{amsmath}
\usepackage{amsfonts}

\usepackage{amsmath,amssymb}
\usepackage{graphicx}
\usepackage{color}
\usepackage{url}

\renewcommand{\arraystretch}{1.3}

\newtheorem{theorem}{Theorem}
\newtheorem{lemma}{Lemma}
\newtheorem{corollary}{Corollary}

\newtheorem{definition}{Definition}
\newtheorem{proposition}{Proposition}
\newtheorem{remark}{Remark}

\newtheorem{example}{Example}

\newcommand{\RNum}[1]{\uppercase\expandafter{\romannumeral #1\relax}}

\newcommand{\gf}{{\mathbb{F}}}

\newcommand{\ls}[1]
    {\dimen0=\fontdimen6\the\font\lineskip=#1\dimen0
     \advance\lineskip.5\fontdimen5\the\font
     \advance\lineskip-\dimen0
     \lineskiplimit=0.9\lineskip
     \baselineskip=\lineskip
     \advance\baselineskip\dimen0
     \normallineskip\lineskip\normallineskiplimit\lineskiplimit
     \normalbaselineskip\baselineskip
     \ignorespaces}

\begin{document}

\bibliographystyle{abbrv}

\title{On the Differential Properties of the Power Mapping $x^{p^m+2}$}

\author{Yuying Man\thanks{Y. Man and Y. Xia are with the Department of Mathematics and Statistics,
South-Central University for Nationalities, Wuhan 430074, China (e-mail:
manyuying122@126.com; xia@mail.scuec.edu.cn). Y. Xia is
  also with the Hubei Key Laboratory of Intelligent Wireless Communications,
  South-Central University for Nationalities, Wuhan 430074, China.}, Yongbo Xia\footnotemark[1] \thanks{Corresponding author},
 Chunlei Li
\thanks{C. Li  and T. Helleseth are with the  Department of Informatics, University of
Bergen, N-5020 Bergen, Norway (e-mail: chunlei.li@uib.no;
tor.helleseth@uib.no).}, and Tor Helleseth \footnotemark[3]}
\date{}
\maketitle

\thispagestyle{plain} \setcounter{page}{1}

\begin{abstract}
Let $m$ be a positive integer and $p$ a prime. In this paper, we investigate the differential properties of the power mapping $x^{p^m+2}$ over $\mathbb{F}_{p^n}$, where $n=2m$ or $n=2m-1$. For the case $n=2m$, by transforming the derivative equation of $x^{p^m+2}$ and studying some related equations, we completely determine the differential spectrum of this power mapping. For the case $n=2m-1$,
the derivative equation can be transformed to a polynomial of degree $p+3$. The problem is more difficult and we obtain partial results about the differential spectrum of $x^{p^m+2}$.

\vspace{2mm}


\end{abstract}

\ls{1.5}

\section{Introduction}\label{sec-intro}

Let $\mathbb{F}_{p^n}$ be the finite field with $p^n$ elements and $\mathbb{F}_{p^n}^*=\gf_{p^n}\setminus \{0\}$, where $p$ is a prime number and $n$ is a positive integer.
Let $F(x)$ be a mapping from  $\mathbb{F}_{p^n}$ to itself. The \textit{derivative function} of $F(x)$ at an element $a$ in $\gf_{p^n}$, denoted by $\mathbb{D}_aF$,  is given  by
$$\mathbb{D}_aF(x)=F(x+a)-F(x).$$
For any $a,\,b \in \gf_{p^n}$,  let  $\delta_F(a,b)=|\{x \in \gf_{p^n}~| ~\mathbb{D}_{a}F(x)=b\}|,$
where $|S|$ denotes the cardinality of a set $S$,
and define
\[\delta(F)=\max \{ \delta_F(a,b)~| ~a \in \mathbb{F}_{p^n}^*,\,\, b \in \mathbb{F}_{p^n}\}.\]
Nyberg defined a mapping $F(x)$ to be \emph{differentially $\delta$-uniform} if and only if $\delta(F)=\delta$ \cite{Nyberg1994}, and $\delta(F)$ is called the \emph{differential uniformity} of $F(x)$ accordingly.
The differential uniformity is an important concept in cryptography since it quantifies the security of the mappings used in many block ciphers. The possible value of the differential uniformity of a mapping over finite fields was discussed in \cite{qu-mathchina-2013}. For practical applications in cryptography, it is usually desirable to employ mappings with differential uniformity no greater than $4$. For example, the AES uses the inverse function $x \mapsto x^{-1}$ over $\gf_{2^n}$,
which has differential uniformity $4$ for even $n$ and $2$ for odd $n$.
Therefore, finding functions with low differential uniformity is of great interest. Much work has been done in this direction. Such results can be found in \cite{Carlet-mathchina-2013,Zeng-mathchina-2013,zhazhengbang-mathchina-2015,Peng-mathchina-2016} and references therein.

Besides the differential uniformity, the differential spectrum of a nonlinear mapping reflects more information about its differential property. The \emph{differential spectrum} of a mapping $F(x)$ of $\mathbb{F}_{p^n}$ to itself is defined as the multiset
\begin{equation*}
\{\,
\delta_F(a,b)\,:\, a \in \gf_{p^n}^*,\, b\in \gf_{p^n}
\,\}.
\end{equation*} It can be used to analyze the resistance of a mapping against some variants of differential cryptanalysis, especially the truncated differential attacks \cite{BCC2010IJICT}.
In addition to its importance in cryptography, the differential spectrum of a nonlinear mapping also plays a significant role in sequences \cite{DHK2011IT,Carlet2018IT,CP2019FFA}, coding theory \cite{CCZ1998DCC,CP2019AMC} and combinatorial design \cite{TangDing2020IT}. One can find more details about these
applications in \cite{YXLIT} and references therein.

When $F(x)$ is  a power mapping, i.e.,  $F(x)=x^d$ for {a positive integer} $d$,  one can easily see that $\delta_F(a,b)=\delta_F(1,{b/{a^d}})$  for all $a\in \gf_{p^n}^*$ and $b\in \gf_{p^n}$.
That is to say,  the differential spectrum of $F(x)$ is completely determined by the values of $\delta_F(1,b)$ as $b$ runs through $\gf_{p^n}$.
Therefore,  the differential spectrum of a power mapping  can be simplified  as follows.

\begin{definition}\label{def1}
Assume that a power function $F(x)=x^d$ over $\gf_{p^n}$ has differential uniformity $\delta$  and denote
\[\omega_i=|\left\{b\in \gf_{p^n}\mid \delta_F(1, b)=i\right\}|,\,\,0\leq i\leq \delta.\]
The differential spectrum of $F(x)$ is simply defined to be
an ordered sequence
\[
\mathbb{S} = [\omega_0, \omega_1, \ldots, \omega_{\delta}].
\]
\end{definition}

For a mapping $F(x)$ from $\mathbb{F}_{p^n}$ to itself, in order to measure the global injectivity and surjectivity of its derivative functions $\mathbb{D}_aF=F(x+a)-F(x)$, $a\in \mathbb{F}_{p^n}^*$, Panario et al. \cite{DSA} introduced the definitions of  ambiguity  and deficiency for $F(x)$, which are closely related to the differential spectrum of $F(x)$.

\begin{definition}\label{def2}
Let $F(x)$ be a mapping over $\gf_{p^n}$ with differential uniformity $\delta$, and define
$$\overline{\omega_i}=|\{ (a,b)\in \mathbb{F}_{p^n}^*\times \mathbb{F}_{p^n} \mid \delta(a,b)=i\}|.$$
The ambiguity of $F(x)$, denoted by $A(F)$,  is defined as
$$A(F)=\sum\limits_{i=2}^{\delta} \overline{\omega}_i\binom{i}{2},$$
and the deficiency  $D(F)$ of $F(x)$ is defined as
$$D(F)=\overline{\omega}_0.$$
\end{definition}

In fact, $A(F)$ is the total replication of pairs $(x,x^{\prime})$ such that $\mathbb{D}_aF(x)=\mathbb{D}_aF(x^{\prime})$ for some $a\in\mathbb{F}_{p^n}^*$, and it is a measure of the injective of the derivative functions $\mathbb{D}_aF$, $a\in \mathbb{F}_{p^n}^*$: the lower the ambiguity of $F(x)$, the closer the functions $\mathbb{D}_aF$ are to injective.  While $D(F)$ is the number of pairs $(a,b)$ such that $\mathbb{D}_aF(x)=b$ has no solution, and it is a measure of the surjective of $\mathbb{D}_aF$, $a\in \mathbb{F}_{p^n}^*$: the lower the deficiency, the closer the functions $\mathbb{D}_aF$ are to surjective. According to Definition \ref{def2}, it is easily seen that knowing the differential spectrum of $F(x)$ can imply its ambiguity and deficiency.

For a nonlinear mapping with low differential uniformity, it is an interesting topic to completely determine its differential spectrum. However, this problem is relatively challenging. A nice survey on this topic can be found in \cite{YXLIT}. We list all the power mappings with known differential spectra in Table \ref{table-1}.

\begin{table}[t]\label{table-1}
\caption{Some power mappings $F(x)=x^d$ over $\gf_{p^n}$ with known  differential spectrum}
\centering

\begin{tabular}{|c|c|c|c|c|}
		\hline
		$p$ & $d$ & Condition & $\delta(F)$ &  Ref.\\  \hline
		$2$ & $2^t+1$ & $\gcd(t, n)=s$ & $2^s$ &\cite{BCC2010IJICT}\\  \hline
		$2$ & $2^{2t}-2^t+1$ & $\gcd(t, n)=s$, $\frac{n}{s}$ odd & $2^s$ &\cite{BCC2010IJICT}\\  \hline
		$2$ & $2^{n}-2$ & $n\ge 2$ & $2$ or $4$ &\cite{BCC2010IJICT}\\ \hline
		$2$ & $2^{2k}+2^k+1$ & $n=4k$ & $4$ &\cite{BCC2010IJICT, XY2017FFA}\\ \hline
		$2$ & $2^t-1$ & $t=3, \, n-2$ & $6$ or $8$ &\cite{BCC2011IT}\\ \hline
		$2$ & $2^t-1$ & $t=\frac{n-1}{2},\,\,\frac{n+3}{2},\,\, n$ odd & $6$ or $8$ &\cite{BP2014DCC}\\  \hline
		$2$ & $2^m+2^{(m+1)/2}+1$ & $n=2m$, $m\geq5$ odd &  $8$ &\cite{XYY2017DCC}\\  \hline
		$2$ & $2^{m+1}+3$ & $n=2m$, $m\geq5$ odd &  $8$ &\cite{XYY2017DCC}\\ \hline
		$2$ & $2^{3k}+2^{2k}+2^k-1$ & $n=4k$ & $2^{2k}$ &\cite{LWZT2020}\\ \hline
		$3$ & $2\cdot3^{(n-1)/2}+1$ & $n$ odd & $4$ &\cite{DHK2011IT}\\  \hline
		$5$ & $\frac{5^n-3}{2}$ & any $n$ & $4$ or $5$ &\cite{YL2021DCC}\\  \hline
		$p$ odd & $p^{2k}-p^k+1$ & $\gcd(n,k)=e$, $\frac{n}{e}$ odd & $p^e+1$ &\cite{YZW2019IJICT, LRF2021AMC}\\  \hline
		$p$ odd & $\frac{p^k+1}{2}$ & $\gcd(n,k)=e$ & $\frac{p^e-1}{2}$ or $p^e+1$ &\cite{CHNC2013FFA}\\  \hline
		$p$ odd & $\frac{p^n+1}{p^m+1}+\frac{p^n-1}{2}$ & $p\equiv 3\,\, ({\rm mod}\,\,4)$, $m|n$, $n$ odd & $\frac{p^m+1}{2}$ &\cite{CHNC2013FFA}\\  \hline
		$p$ odd & $p^n-3$ & any $n$ & $\le 5$ &\cite{XZLH2020FFA,YXLIT}\\  \hline
		$p$ odd & $p^m+2$ & $p>3,\,\,n=2m$ & $4$ &This paper\\  \hline	
\end{tabular}
\end{table}

In the present paper, we will investigate the differential spectrum of the power mapping $x^{p^m+2}$ over $\mathbb{F}_{p^n}$, where $p$ is a prime, $m$ is a positive integer, $n=2m$ or $n=2m-1$. When $p\ge 3$ and $n=2m$, Helleseth, Rong and Sandberg studied the differential uniformity of this mapping in \cite{HRS1999IT}, where they showed that the differential uniformity of $x^{p^m+2}$ is equal to $2$ if $p^m\equiv1\,\, (\mbox{mod} \,\,3)$, and equal to $4$ if $p^m\equiv2\,\, (\mbox{mod} \,\,3)$. By carefully checking, we find that their proof in \cite{HRS1999IT}, to some extent, is not sufficient, and more explanations should be added. Moreover, if one wants to determine the differential spectrum of this mapping, some new techniques should be proposed to investigate the derivative equation of $x^{p^m+2}$.

The remainder of this paper is organized as follows.  Section \ref{pre}  gives some preliminary results that will be needed in this paper. The result for the case $n=2m$ is presented in Section \ref{main-result}. Section \ref{result-in-odd} deals with the case $n=2m-1$, and the concluding remarks are given in Section \ref{con-remarks}.

\section{Preliminaries}\label{pre}
In order to prove our main result in this paper, we need to make some preparations. Let $F(x)=x^d$ be a differentially $\delta$-uniform power mapping over $\gf_{p^n}$. According to Definition \ref{def1}, we have the following identities
\begin{equation}\label{prop}
\sum\limits_{i=0}^\delta \omega_i=p^n\,\,{\rm and}\,\,\sum\limits_{i=0}^\delta \left(i\times \omega_i\right)=p^n,
\end{equation}
which are very useful in computing the differential spectrum of $F(x)$.

Let $q$ be a power of a prime, and $\mathbb{F}_q[x]$ be the polynomial ring over $\mathbb{F}_q$. The following lemmas will be used frequently in this paper.

\begin{lemma}\label{lemma1-quad}(\cite{XZLH2020FFA})
	The polynomial $Q(x)=x^2+ax+b\in \mathbb{F}_q[x]$, $q$ odd, is irreducible in $\mathbb{F}_q[x]$ if and only if $a^2-4b$ is a nonsquare in $\mathbb{F}_q$. In particular, if $a^2-4b$ is a nonzero square in $\mathbb{F}_q$, then $Q(x)$ has two distinct roots in $\mathbb{F}_q$.
\end{lemma}

When $q$ is odd, for a square element $s\in \gf_{q}$, it has exactly two square roots in $\mathbb{F}_q$. We will use $\pm \sqrt{s}$ to express the two square roots of $s$ throughout this paper. The following lemma is about the number of roots to a special linearized polynomial over $\gf_{p^n}$. For the reader's
convenience, we include a proof here.
\begin{lemma}\label{lemma2-equation}
Let $p$ be a prime, $n$ and $t$ be two positive integers, and $s=\gcd(n, t)$. Then, the linearized polynomial $x^{p^t}+x$ has $p^s$ roots in $\gf_{p^n}$ if $p=2$, and $p^{\frac{1+(-1)^{n/s}}{2}s}$ roots if $p$ is odd.
\end{lemma}
{\em Proof:}
If $p=2$, then we have $x^{2^t}=x$, which implies that $x\in \gf_{2^t}$. Since $x\in \gf_{2^n}$, it then follows that $x\in \gf_{2^s}$. Thus in this case $x^{2^t}+x$ has $2^s$ roots in  $\gf_{2^n}$.

If $p>2$, then the nonzero roots of  $x^{p^t}+x$ satisfy $x^{p^t-1}=-1$. Let $\alpha$ be a primitive element of $\gf_{p^n}$ and assume that $x=\alpha^i$. Then the equation $x^{p^t-1}=-1$ is equivalent to the congruence  $i(p^t-1)\equiv \frac{p^n-1}{2}\,\, ({\rm mod}\,\, p^n-1)$.  This congruence has solutions if and only if  $\left(p^s-1\right)|\frac{p^n-1}{2}$, and when it has solutions, the number of its solutions is equal to $p^s-1$. Note that $\frac{p^n-1}{2}=\frac{p^{us}-1}{2}=\frac{(p^s-1)(p^{s(u-1)}+p^{s(u-2)}+\cdots +p^s+1)}{2}$, where $u=n/s$. To ensure that $\left(p^s-1\right)|\frac{p^n-1}{2}$, $\frac{p^{s(u-1)}+\cdots +p^s+1}{2}$ must be an integer, which implies $u$ must be even. Therefore, $x^{p^t-1}=-1$ has $p^s-1$ solutions in $\gf_{p^n}$ if $u$ is even, and has no solution if $u$ is odd.

Summarizing the above discussions, we obtain the desired result.
\hfill$\square$

\begin{lemma}\label{quadartic-3}
Let $p>3$ be a prime and $m$ be a positive integer. Then, $-3$ is a nonsquare in $\gf_{p^m}$ if and only if  $p^m\equiv 2\,\, ({\rm mod}\,\,3)$, and a square if and only if  $p^m\equiv 1 \,\,({\rm mod}\,\,3)$.
\end{lemma}
{\em Proof:}  Note that since $p>3$ is a prime, either $p^m\equiv 1\,\, ({\rm mod}\,\,3)$ or $p^m\equiv 2\,\, ({\rm mod}\,\,3)$. Thus, we only need to prove that $-3$ is a nonsquare in $\gf_{p^m}$ if and only if  $p^m\equiv 2\,\, ({\rm mod}\,\,3)$.
It is easily seen that if $-3$ is a nonsquare in $\gf_{p^m}$, then it is a nonsquare in $\gf_p$ and $m$ is odd. Let $\left(\frac{\cdot}{p}\right)$ denote the Legendre symbol
from elementary number theory.  By the law of quadratic reciprocity \cite{Pan-num-theory}, we have
$$\left(\frac{3}{p}\right)=(-1)^{\frac{p-1}{2}}\left(\frac{p}{3}\right).$$
Then, we have
\begin{equation}\label{relation-for-p-3}
\begin{array}{lcl}
\left(\frac{-3}{p}\right)&=&\left(\frac{-1}{p}\right)\left(\frac{3}{p}\right)\\
&=&(-1)^{\frac{p-1}{2}}\left(\frac{3}{p}\right)\\
&=&(-1)^{\frac{p-1}{2}}(-1)^{\frac{p-1}{2}}\left(\frac{p}{3}\right)\\
&=&\left(\frac{p}{3}\right).\\
\end{array}
\end{equation}
Since $-3$ is a nonsquare in $\gf_{p}$, we have  $\left(\frac{-3}{p}\right)=-1$, which implies $p\equiv 2\,\, ({\rm mod}\,\,3)$ by (\ref{relation-for-p-3}).  Thus, we have $p^m\equiv 2\,\, ({\rm mod}\,\,3)$ since $m$ is odd.

Conversely, suppose that $p^m\equiv 2\,\, ({\rm mod}\,\,3)$. One immediately concludes that  $p\equiv 2\,\, ({\rm mod}\,\,3)$  and $m$ is odd. Then, by (\ref{relation-for-p-3}), we have  $\left(\frac{-3}{p}\right)=\left(\frac{p}{3}\right)=\left(\frac{2}{3}\right)=-1$. This shows that in this case $-3$ is a nonsquare in $\gf_p$, and  so it is a nonsquare in $\gf_{p^m}$  since $m$ is odd. \hfill$\square$

\section{The main result for the case $n=2m$}\label{main-result}
The  aim of this section is to determine the differential spectrum of the power mapping  \begin{equation}\label{mapping-studied}F(x)=x^{p^m+2}\end{equation} over $\gf_{p^n}$, where $p$ is a prime and $n=2m$ with $m$ being a positive integer. For $p>3$, Hellesth, Rong and Sndberg
studied the differential uniformity of this mapping in \cite{HRS1999IT}, where they presented the possible values of the differential uniformity. From their proof, one cannot find enough information to derive the differential spectrum of $F(x)$. In order to compute the differential spectrum, the derivative equation $\mathbb{D}_1F(x)=F(x+1)-F(x)=b$ should be further investigated.

Next we will consider the derivative equation of $F(x)$, which is given below
\begin{equation}\label{diff-equa}
\mathbb{D}_1F(x)=(x+1)^{p^m+2}-x^{p^m+2}=b,
\end{equation}
where $b \in \gf_{p^n}$.  For simplicity, we use $\delta(b)$ instead of $\delta_F(1,b)$ to denote the number of solutions of (\ref {diff-equa}) in $\gf_{p^n}$. According to Definition \ref{def1}, the problem of calculating the differential spectrum of $F(x)$ can be reduced to that of determining the value distribution of $\delta(b)$ as $b$ runs through $\gf_{p^n}$.  It is easily seen that (\ref{diff-equa}) can be written as
\begin{equation*}
2x^{p^m+1}+x^{p^m}+x^2+2x+1-b=0.
\end{equation*}
Throughout this section, we will use $\bar{x}$ to denote  $x^{p^m}$. Let $a=b-1$, then above equation becomes
\begin{equation}\label{diff-equa2}
2x\bar{x}+\bar{x}+x^2+2x-a=0.	
\end{equation}

When $p=2$, (\ref{diff-equa2}) is equivalent to
\begin{equation}\label{diff-2^m+2}
\bar{x}+x^2=a,	
\end{equation}
which is an affine linearized polynomial over $\gf_{p^n}$ and its number of solutions can be easily derived.
When $p>2$, with the linear transformation $x=y-\frac{1}{2}$, (\ref{diff-equa2}) becomes
\begin{equation}\label{diff-equa3}
8y\bar{y}+4y^2-c=0,
\end{equation}
where $c=4a+3$.
If $c\neq 0$, then $y\neq 0$ and  from (\ref{diff-equa3}) we further have
\begin{equation}\label{ybar}
\bar{y}=\frac{-y^2+\frac{c}{4}}{2y}.
\end{equation}
Raising (\ref {diff-equa3}) to the  power $p^m$, we obtain
\begin{equation}\label{diff-equa4}
8\bar{y}y+4\bar{y}^2-\bar{c}=0.
\end{equation}
Substituting (\ref{ybar}) into (\ref {diff-equa4}), we get
\begin{equation}\label{diff-equa5}
3y^4+\frac{2\bar{c}-c}{2}y^2-\frac{c^2}{16}=0,
\end{equation}
which becomes
\begin{equation}\label{diff-3^m+2}
(\bar{a}+a)y^2-a^2=0
\end{equation} in the case $p=3$. Thus, in the sequel we should distinguish three cases: $p=2$, $p=3$ and $p>3$.

\subsection{The differential spectrum of $x^{p^m+2}$ for $p=2$}\label{p=2}
When $p=2$, the derivative equation  of  $F(x)$ is given in (\ref{diff-2^m+2}). We first consider the
associated linearized polynomial $L(x)=\bar{x}+x^2$, which is a linear transformation
from $\mathbb{F}_{2^n}$ to $\mathbb{F}_{2^n}$.  Let $ {\rm Im}(L)$ denote the image of $L(x)$, and ${\rm Ker}(L)$ be the kernel of $L(x)$.
Note that $L(x)=\bar{x}+x^2=0$ if and only if $x^{2^{m-1}}+x=0$. By Lemma \ref{lemma2-equation}, the latter equation has $2^{\gcd(m-1,n)}$ solutions in $\gf_{2^n}$. Thus, the linearized polynomial $L(x)=0$ also has $2^{\gcd(m-1,n)}$ solutions. This implies that $|{\rm Ker}(L)|=2^{\gcd(m-1,n)}=2^{\gcd(m-1,2)}$. By the homomorphism theorem for groups, we also have $|{\rm Im}(L)|=2^{n-\gcd(m-1,2)}$.

For the derivative equation $L(x)=\bar{x}+x^2=a$ in (\ref{diff-2^m+2}), it has  $2^{\gcd(m-1,2)}$ solutions if $ a\in {\rm Im}(L)$ and no solution otherwise.  Thus, there are exactly $|{\rm Im}(L)|$ $b$'s in $\gf_{2^n}$ such that $\delta(b)=2^{\gcd(m-1,2)}$, and $(2^n-|{\rm Im}(L)|)$ $b$'s such that $\delta(b)=0$ since $a=b-1$. Note that $\gcd(m-1,2)=1$ or $2$. Then, we can obtain the following result.

\begin{theorem}\label{dsforp=2}Let $F(x)$ be the power mapping over $\gf_{p^n}$ defined as in (\ref{mapping-studied}). When $p=2$, its differential spectrum is given by
$$[\omega_0=2^{n-1},\omega_1=0,\omega_2=2^{n-1}]$$
for even $m$, and $$[\omega_0=2^n-2^{n-2}, \omega_1=0, \omega_2=0, \omega_3=0, \omega_4=2^{n-2}]$$ for odd $m$.
\end{theorem}

\begin{remark}\label{rem1}
Blondeau et al. studied the differential spectrum of the Gold function $x^{2^t+1}$ over $\gf_{2^n}$ \cite{BCC2010IJICT}. Since $x^{2^m+2}=x^{2(2^{m-1}+1)}$, it is clear that the power mapping $x^{2^m+2}$ over $\gf_{2^n}$ has the same differential spectrum as that of the Gold-type function $x^{2^{m-1}+1}$. Thus, we can also derive the result in Theorem \ref{dsforp=2} from \cite{BCC2010IJICT}. Here we give the proof to make our paper self-contained. Moreover, by Lemma \ref{lemma2-equation}, one can easily compute the differential spectrum of the power mapping $x^{p^t+1}$ over $\gf_{p^n}$ for any odd prime $p$.
\end{remark}

\subsection{The differential spectrum of $x^{p^m+2}$ for $p=3$}
In this subsection, we will determine the differential spectrum of $F(x)$ for the case $p=3$. Note that in this case the exponent $p^m+2$ is of  Niho-type \cite{XIA2017IT}.  The main result is presented in the following theorem.

\begin{theorem}\label{dsforp=3}Let $F(x)$ be the power mapping over $\gf_{p^n}$ defined as in (\ref{mapping-studied}), and $\delta(b)$ denote the number of solutions of the derivative equation $\mathbb{D}_1F(x)=F(x+1)-F(x)=b$. When $p=3$, we have that $\delta(1)=3^m$ and $\delta(b)\in \{0,2\}$ for any $b\in \gf_{p^n}\setminus \{1\}$. Moreover, the differential spectrum  of $F(x)$ in this case is given by
$$[\omega_0=\frac{3^n+3^m}{2}-1,\omega_1=0,\omega_2=\frac{3^n-3^m}{2},\omega_3=0,\cdots,\omega_{3^m-1}=0,\omega_{3^m}=1].$$
\end{theorem}
{\em Proof:}  When $p$ is odd, the derivative equation  (\ref{diff-equa}) is equivalent to  (\ref{diff-equa3}), which can be further rewritten as
\begin{equation}\label{specfor3}
-\bar{y}y+{\bar{y}}^2-\bar{a}=0
\end{equation}
in the case $p=3$, where $a=b-1$. If $b=1$, then the above equation becomes $-y\bar{y}+y^2=0$, which implies $\bar{y}=y$, i.e., $y \in \gf_{3^m}$. This shows that $\delta(1)=3^m$.

Next we consider the case $b\neq 1$, i.e., $a\neq 0$. Then by the discussions prior to Subsection \ref{p=2} the solutions of (\ref{specfor3}) must satisfy (\ref{diff-3^m+2}). Note that for given $a\neq0$, (\ref{diff-3^m+2}) has $0$ or $2$ solutions in $\gf_{3^n}$. Thus, in this case (\ref{specfor3}) has at most two solutions. If $y_0$ is a solution of (\ref{specfor3}), then $y_0\neq 0$ since $a\neq 0$ and $-y_0$ is also a solution of (\ref{specfor3}). This shows that the solutions of (\ref{specfor3}) come in pairs. Hence, in this case the possible numbers of solutions of (\ref{specfor3}) are $0$ and $2$. So we have $\delta(b)\in\{0,2\}$ for $b\neq 1$.

The above discussions imply that in the differential spectrum of $F(x)=x^{3^m+2}$, the component $\omega_i$ is equal to $0$ if $i\notin\{ 0,2,3^m\}$. Since $\omega_{3^m}$ has been already known, we now only need to determine $\omega_0$ and $\omega_2$.
According to the identities in (\ref{prop}), we get
\begin{equation*}\arraycolsep=1.2pt\def\arraystretch{1.7}
\left \{\begin{array}{lll}
\omega_0+\omega_2+\omega_{3^m}=p^n,\\
2\omega_2+3^m\omega_{3^m}=p^n,\\
\omega_{3^m}=1.
\end{array}\right.\end{equation*}
Solving this equation system, the differential spectrum of $F(x)$ for $p=3$ is obtained.
             \hfill$\square$

\subsection{The differential spectrum of $x^{p^m+2}$ for $p>3$}
In this subsection the differential spectrum of $F(x)$ for $p>3$ will be determined.
With the notation in Theorem \ref{dsforp=3}, we have the following result.
\begin{lemma}\label{lemma-num} When $p>3$, for each given $b\in\gf_{p^n}$, $\delta(b)\in\{0,1,2,4\}$. Especially, $\delta(b)=1$ if and only if $b=\frac{1}{4}$.
\end{lemma}
{\em Proof:}  According to the discussions presented at the beginning of Section \ref{main-result}, when $p>3$, the derivative equation $\mathbb{D}_1F(x)=b$ in (\ref {diff-equa}) is equivalent to (\ref {diff-equa3}), where $c=4a+3=4b-1$ and $y=x-\frac{1}{2}$.  If $c=0$, i.e., $b=\frac{1}{4}$, then (\ref {diff-equa3}) becomes $8y\bar{y}+4y^2=0$, which leads to $y=0$ or $2\bar{y}+y=0$. For the latter case, it is equivalent to $2y^{p^m-1}+1=0$ if $y\neq 0$. Then, one gets $y^{p^m-1}=-\frac{1}{2}$. By raising this
identity to the power ($p^m+1$), we get $(y^{p^m-1})^{p^m+1}=(-\frac{1}{2})^{p^m+1}$, which leads to $1=\frac{1}{4}$, a contradiction. Thus, if $c=0$, then (\ref {diff-equa3}) has exactly one solution $y=0$.

 Next we consider that case $c\neq 0$. In this case the solutions of (\ref {diff-equa3}) are also solutions of (\ref{diff-equa5}). Note that (\ref{diff-equa5}) has at most four solutions in $\gf_{p^n}$ since it is of degree four. Thus, when $c\neq 0$, (\ref {diff-equa3}) has at most four solutions as well. On the other hand, if $y_0$ is a solution of (\ref {diff-equa3}), then $-y_0$ is also a solution of (\ref {diff-equa3}). When $c\neq 0$, $y_0\neq -y_0$ and so the solutions of (\ref {diff-equa3}) come in pairs. This shows that when $c\neq 0$, the possible numbers of  solutions of (\ref {diff-equa3}) are $0$, $2$ and $4$.

Based on the above discussion and the relationship between (\ref {diff-equa}) and (\ref {diff-equa3}),  the desired result follows.
\hfill$\square$

According to Lemma \ref{lemma-num},  when $p>3$ the differential uniformity $\delta(F)$ of $F(x)$ is at most $4$. In order to determine the differential spectrum of $F(x)$, we need to count the numbers of $b$ such that (\ref {diff-equa}) has $0$, $2$ and $4$ solutions, respectively. To do this,  we should
further investigate the equations in (\ref {diff-equa3}) and  (\ref{diff-equa5}).  The following lemma will be needed.

\begin{lemma}\label{charcD}
Let $p>3$, $n=2m$, and $\alpha$  a primitive element of $\gf_{p^n}$. Assume that $D=\bar{c}^2-\bar{c}c+c^2$ with  $c\in \mathbb{F}_{p^n}^*$.  Then $D$ satisfies the following properties:

\noindent (i) when $p^m\equiv 1\,\, ({\rm mod}\,\,3)$, $D\neq 0$ for any $c\in \mathbb{F}_{p^n}^*$;

\noindent (ii) when $p^m\equiv 2\,\, ({\rm mod}\,\,3)$, $D=0$ if and only if  $c=\alpha^{\frac{p^m+1}{6}+j(p^m+1)}$ or $c=\alpha^{\frac{5(p^m+1)}{6}+j(p^m+1)}$ with $j\in \{0, 1, \cdots, p^m-2\}$.
\end{lemma} 	
{\em Proof:}  Define
\begin{equation*}
\mathcal{U}=\left\{ x\,|\,x^{p^m+1}=1,\,\, x\in \gf_{p^n}\right\}.
\end{equation*}
Then $\mathcal{U}$ is a cyclic subgroup of order $p^m+1$ in the multiplicative group $\gf_{p^n}^*$ with
generator $\beta=\alpha^{p^m-1}$, and it can be expressed as
\begin{equation*}
\mathcal{U}=<\beta>=\left\{\alpha^0, \alpha^{p^m-1}, \alpha^{2(p^m-1)}, \cdots, \alpha^{p^m(p^m-1)}\right\}.
\end{equation*}
 For any $c\in \mathbb{F}_{p^n}^*$, $D=\bar{c}^2-\bar{c}c+c^2=0$ implies that  $(\frac{\bar{c}}{c})^2-(\frac{\bar{c}}{c})+1=0$. The latter equation is equivalent to $\left(\frac{\bar{c}}{c}\right)^3+1=0$ with $\frac{\bar{c}}{c}\neq -1$. Note that $\alpha^{\frac{p^{2m}-1}{2}}=\beta^{\frac{{p^m+1}}{2}}=-1$ and $\frac{\bar{c}}{c}\in \mathcal{U}$. Let $u=\frac{\bar{c}}{c}$ and suppose that  $u=\beta^i$ for some $i\in \{0, 1, \cdots, p^m\}$. Then, $u^3+1=0$ is equivalent to the following congruence
\begin{equation}\label{mod}
3i\equiv \frac{p^m+1}{2}\,\, ({\rm mod}\,\, p^m+1).
\end{equation}
Since $\gcd(3,p^m+1)=\gcd(3,\frac{p^m+1}{2})\mid \frac{p^m+1}{2}$, the congruence (\ref{mod}) has $\gcd(3,p^m+1)$ solutions in $\mathbb{Z}_{p^m+1}$. In order to exclude the case $u=-1$, the solution $i=\frac{p^m+1}{2}$ of (\ref{mod}) should be removed.

When $\gcd(3,p^m+1)=1$, i.e., $p^m\equiv 1\,\, ({\rm mod}\,\,3)$, (\ref{mod}) has exactly one solution $i=\frac{p^m+1}{2}$ in $\mathbb{Z}_{p^m+1}$. So in this case there is no $c\in\gf_{p^n}^*$ such that $D=0$. This means when $p^m\equiv 1\,\, ({\rm mod}\,\,3)$, $D\neq 0$ for any $c\in\gf_{p^n}^*$.

When $p^m\equiv 2\,\, ({\rm mod}\,\,3)$, $\gcd(3,p^m+1)=3$. In this case, besides $i=\frac{p^m+1}{2}$, (\ref{mod}) also has other two solutions in $\mathbb{Z}_{p^m+1}$: $i=\frac{p^m+1}{6}$ and $i=\frac{5(p^m+1)}{6}$. Thus, we get $u=\beta^{\frac{p^m+1}{6}}$ or $u=\beta^{\frac{5(p^m+1)}{6}}$. Let $c=\alpha^t$ with $t\in \{0, 1, \cdots, p^n-2\}$. Then, the problem of finding the solutions of $D=0$ in this case is identical with that of obtaining all solutions of
\begin{equation*}
t(p^m-1)\equiv (p^m-1)\frac{p^m+1}{6}\,\, ({\rm mod}\,\, p^n-1)
\end{equation*}
and \begin{equation*}
t(p^m-1)\equiv (p^m-1)\frac{5(p^m+1)}{6}\,\, ({\rm mod}\,\, p^n-1).
\end{equation*}
Solving the above congruences, we obtain $t=\frac{p^m+1}{6}+j(p^m+1)$ or $t=\frac{5(p^m+1)}{6}+j(p^m+1)$ with $j\in \{0, 1, \cdots, p^m-2\}$. The desired result then follows immediately.
	\hfill$\square$	

Recall that the derivative equation $\mathbb{D}_1F(x)=b$ has been transformed to
	$8y\bar{y} + 4y^2 -c=0$ in (\ref{diff-equa3}), which can induce the quartic equation
$3y^4+\frac{2\bar{c}-c}{2}y^2-\frac{c^2}{16}=0$ in (\ref{diff-equa5}) when $c\neq0$. 
 The following lemma gives some elaborate properties about the equation (\ref{diff-equa3}).

\begin{lemma}\label{lemma5D}With the notation introduced above, we have the following results:

\noindent (i) when $p^m\equiv 2\,\, ({\rm mod} \,\,3)$, for each $c\in \mathbb{F}_{p^n}^*$ such that $D\neq 0$, the equation (\ref{diff-equa3}) cannot have two solutions in $\gf_{p^n}$;

\noindent (ii) when $p^m\equiv 1\,\, ({\rm mod} \,\,3)$, for any $c\in \mathbb{F}_{p^n}$, the equation (\ref{diff-equa3}) cannot have four solutions.
\end{lemma}	
{\em Proof:} By Lemma \ref{lemma-num}, when $c=0$, the equation (\ref{diff-equa3}) has exactly one solution in $\mathbb{F}_{p^n}$, which is $y=0$. When $c\neq0$,  (\ref{diff-equa3}) have $0$, $2$ or $4$ solutions in $\mathbb{F}_{p^n}$, and all its solutions satisfy (\ref{diff-equa5}). Thus, we first investigate the solutions of (\ref{diff-equa5}) under the condition that $c\ne 0$.  Let $z=y^2$, then (\ref{diff-equa5}) becomes
\begin{equation}\label{equa5-equiv}
3z^2+\frac{2\bar{c}-c}{2}z-\frac{c^2}{16}=0,	
\end{equation} which is a quadratic equation in variable $z$. The discriminant of (\ref{equa5-equiv}) is $\frac{D}{9}$. It is obvious that $D\in\mathbb{F}_{p^m}$ since $D=\bar{D}$. If $D\neq 0$, then the discriminant is a nonzero square in $\gf_{p^n}$ and (\ref{equa5-equiv}) has exactly two distinct solutions $z_1$, $z_2$ in $\gf_{p^n}^*$ due to Lemma \ref{lemma1-quad}. Thus, we can rewrite  (\ref{equa5-equiv}) as $(z-z_1)(z-z_2)=0$, where $z_1$ and $z_2$ satisfy
\begin{equation}\label{z-relation}
\left\{\begin{array}{ll}
z_1+z_2=-\frac{2\bar{c}-c}{6},\\
z_1z_2=-\frac{c^2}{48}.
\end{array}\right.\end{equation}
Note that $-\frac{c^2}{48}$ is always a square in $\mathbb{F}_{p^n}^*$. Thus, $z_1$ and $z_2$ are either both squares in $\mathbb{F}_{p^n}^*$, or both nonsquares in $\mathbb{F}_{p^n}^*$.

 Now we suppose that $z_1$ and $z_2$ are both squares in $\gf_{p^n}^*$, that is, $z_i=y_i^2$ for some $y_i\in\gf_{p^n}^*$, where $i\in\{1,2\}$. Then, (\ref{diff-equa5}) has four solutions in $\gf_{p^n}^*$, which are $\pm y_i$, $i=1,2$. If $y_i$ is a solution of (\ref{diff-equa3}), so is $-y_i$. Next we should decide whether $y_i$ satisfies (\ref{diff-equa3}), $i=1, 2$. To this end, we compute $8y_1\bar{y_1} \cdot 8y_2\bar{y_2}$ as follows
\begin{equation*}
8y_1\bar{y_1} \cdot 8y_2\bar{y_2}=64(y_1^2y_2^2)^{\frac{p^m+1}{2}}=64(z_1z_2)^{\frac{p^m+1}{2}},
\end{equation*}
which together with (\ref{z-relation}) implies that
\begin{equation}\label{vleft}
8y_1\bar{y_1} \cdot 8y_2\bar{y_2}=64(-\frac{c^2}{48})^{\frac{p^m+1}{2}}=-\frac{4}{3}(-\frac{1}{3})^{\frac{p^m-1}{2}}\cdot \bar{c}c.
\end{equation}
On the other hand, we compute
\begin{equation}\label{vright}
(c-4y_1^2)(c-4y_2^2)=16y_1^2y_2^2-4c(y_1^2+y_2^2)+c^2=16(z_1z_2)-4c(z_1+z_2)+c^2=\frac{4}{3}\bar{c}c.
\end{equation}
 Note that $(-\frac{1}{3})^{\frac{p^m-1}{2}}=(-3)^{\frac{p^m-1}{2}}$, and it is equal to $1$ or $-1$ depending on whether $-3$ is a square in $\gf_{p^m}$ or not. Next we consider the following two cases.

{\it \textbf{Case 1:}} $p^m\equiv 2\,\, ({\rm mod}\,\, 3)$. This means $-3$ is a nonsquare in $\gf_{p^m}$ due to Lemma  \ref{quadartic-3}. Then $(-\frac{1}{3})^{\frac{p^m-1}{2}}=-1$. In this case, we assume that for $c\in\mathbb{F}_{p^n}^*$ such that $D\neq 0$, (\ref{diff-equa3}) has exactly two solutions. Accordingly, the corresponding equation (\ref{equa5-equiv}) has two distinct solutions $z_1$ and $z_2$, which are both squares in $\mathbb{F}_{p^n}^*$. Furthermore, from  (\ref{vleft}) and (\ref{vright}) we obtain
\begin{equation}\label{case1eq}
8y_1\bar{y_1} \cdot 8y_2\bar{y_2}=\frac{4}{3}\bar{c}c=(c-4y_1^2)(c-4y_2^2).
\end{equation}
On the other hand, since (\ref{diff-equa3}) has exactly two solutions, one can conclude that one and only one of $\{y_1,y_2\}$ satisfies (\ref{diff-equa3}). It then follows that $8y_1\bar{y_1} \cdot 8y_2\bar{y_2}\neq(c-4y_1^2)(c-4y_2^2)$, which contradicts (\ref{case1eq}). Therefore, under the given conditions,  (\ref{diff-equa3}) cannot have two solutions.

{\it \textbf{Case 2:}} $p^m\equiv 1\,\, ({\rm mod}\,\, 3)$. This holds if and only if $-3$ is a square in $\gf_{p^m}$.  Then we get $(-\frac{1}{3})^{\frac{p^m-1}{2}}=1$.
In this case, $D\neq 0$ for any $c\in \gf_{p^n}^*$ due to Lemma \ref{charcD}. We have already known that (\ref{equa5-equiv}) has exactly one solution if $c=0$. Now we assume that for $c\neq 0$, (\ref{diff-equa3}) has four solutions. Then, (\ref{equa5-equiv}) has two solutions $z_1$ and $z_2$ that are both squares in $\mathbb{F}_{p^n}^*$. Similarly, from  (\ref{vleft}) and (\ref{vright}), we have
\begin{equation}\label{case2eq}
8y_1\bar{y_1} \cdot 8y_2\bar{y_2}=-\frac{4}{3}\bar{c}c=-(c-4y_1^2)(c-4y_2^2).
\end{equation}	
Since (\ref{diff-equa3}) has four solutions, it then follows that $y_1$ and $y_2$ both satisfy (\ref{diff-equa3}), which implies \begin{equation*}
8y_1\bar{y_1} \cdot 8y_2\bar{y_2}=(c-4y_1^2)(c-4y_2^2),
\end{equation*}	
a contradiction to (\ref{case2eq}).
Thus, in this case (\ref{diff-equa3}) cannot have four solutions.
\hfill$\square$

\begin{remark}\label{two-four-con}
According to the Lemmas \ref{lemma-num} and \ref{lemma5D},  we can conclude that

 \noindent (\textrm{i}) when $p^m\equiv 1\,\, ({\rm mod}\,\, 3)$, the differential uniformity of $F(x)=x^{p^m+2}$ is at most $2$;

  \noindent (\textrm{ii}) when $p^m\equiv 2\,\, ({\rm mod}\,\, 3)$, the equation (\ref{diff-equa3}) can have two solutions only for $c\in\mathbb{F}_{p^n}^*$ that makes $D=0$.
\end{remark}

Now we  assume that the differential spectrum $\mathbb{S}$ of $F(x)$ is given by
\begin{equation}\label{dsforoddp}
\mathbb{S}=[\omega_0,\omega_1,\cdots,\omega_4].
\end{equation}
By Lemma \ref{lemma-num}, we have $\omega_1=1$ and $\omega_3=0$. Moreover, when $p^m\equiv 1\,\, ({\rm mod}\,\, 3)$, we also have $\omega_4=0$  by Remark \ref{two-four-con} (i). Therefore, for the case $p^m\equiv 1\,\, ({\rm mod}\,\, 3)$, based on the identities (\ref{prop}), the differential spectrum of $F(x)$ can be computed. However, for the case $p^m\equiv 2\,\, ({\rm mod}\,\, 3)$ we need more conditions to compute the differential spectrum. Next we will deal with this problem by determining the value of $\omega_2$.

In what follows, we always assume that $p^m\equiv 2\,\, ({\rm mod}\,\, 3)$ unless otherwise stated. Let $\alpha$ be a primitive element of $\gf_{p^n}$ and define
 \begin{equation}\label{defofD}
 \mathcal{D}=\{\alpha^t\,|\, t=\frac{p^m+1}{6}+j(p^m+1)\,\,{\rm  or }\,\,t=\frac{5(p^m+1)}{6}+j(p^m+1)\,\,{\rm  with }\,\,j=0, 1, \cdots, p^m-2\}.
 \end{equation} By Lemma \ref{charcD}, for $c\in\gf_{p^n}^*$, $D=\bar{c}^2-\bar{c}c+c^2=0$  if and only if $c\in\mathcal{D}$.  According to Remark \ref{two-four-con} (ii), the value of $\omega_2$ is equal to the number of  $c$ in $\mathcal{D}$ such that (\ref{diff-equa3}) have exactly two solutions in $\gf_{p^n}$.

 For any $c\in\mathcal{D}$, all the solutions of (\ref{diff-equa3}) satisfy (\ref{diff-equa5}). Since $D=0$, (\ref{diff-equa5}) can be further simplified as $(y^2-(\frac{-2\bar{c}+c}{12}))^2=0$, that is, $y^2=\frac{-2\bar{c}+c}{12}$.
 Substituting $y^2=\frac{-2\bar{c}+c}{12}$ into (\ref{diff-equa3}), one gets $\bar{y}y=\frac{\bar{c}+c}{12}$. Thus, for any $c\in\mathcal{D}$,
 the equation (\ref{diff-equa3}) is equivalent to
 \begin{equation}\label{det-omega2}
\left \{\begin{array}{ll}
y^2=\frac{-2\bar{c}+c}{12},\\
\bar{y}y=\frac{\bar{c}+c}{12},
\end{array}\right.
\end{equation}
and now $\omega_2$ is equal to the number of $c\in\mathcal{D}$ such that (\ref{det-omega2}) has exactly two solutions in $\gf_{p^n}$.

To make sure (\ref{det-omega2}) holds,  we should first verify that $\frac{-2\bar{c}+c}{12}$ is a square in $\gf_{p^n}$. Since the element $12$ is already a square in $\gf_{p^n}$, we only need to check that $-2\bar{c}+c$ is a square in $\gf_{p^n}$. The result is given as follows.

\begin{lemma}\label{lemmasquare}
With the same notation as in Lemma \ref{charcD}, assume that  $p^m\equiv 2\,\, ({\rm mod}\,\, 3)$ and $\mathcal{D}$ is the set defined as in (\ref{defofD}). Then $-2\bar{c}+c$ is always a square in $\gf_{p^n}$ for any $c\in \mathcal{D}$.
\end{lemma}	
{\em Proof:}
When $p^m\equiv 2\,\, ({\rm mod}\,\, 3)$, $-3$ is a nonsquare in $\gf_{p^m}$ due to Lemma \ref{quadartic-3}, and thus $(-3)^{\frac{p^m-1}{2}}\\=-1$. Note that for any $c\in\mathcal{D}$, we have $(-2\bar{c}+c)^2=-3c^2$.  Then,  $-2\bar{c}+c$ is a square in $\gf_{p^n}$ if and only if $-3c^2$ is a fourth power in $\gf_{p^n}$, that is, if and only if  $(-3c^2)^{\frac{p^{2m}-1}{4}}=1$. Note that $(-3c^2)^{\frac{p^{2m}-1}{4}}=(-3)^{\frac{p^m-1}{2}\frac{p^m+1}{2}}\cdot c^{\frac{p^n-1}{2}}=(-1)^{\frac{p^m+1}{2}}\cdot c^{\frac{p^n-1}{2}}$. We should distinguish two cases accordingly.
		
 \emph{\textbf{Case 1:}} $p^m\equiv 1\,\, ({\rm mod}\,\, 4)$. Then $\frac{p^m+1}{2}$ is odd and thus $(-1)^{\frac{p^m+1}{2}}=-1$.  Therefore, for any 
 $c\in\mathcal{D}$, $-2\bar{c}+c$ is a square in $\gf_{p^n}$ if and only if $c^{\frac{p^n-1}{2}}=-1$, that is, if and only if $c$ is a nonsquare in $\gf_{p^n}$. By the definition of $\mathcal{D}$ in (\ref{defofD}), it can be easily checked that when $p^m\equiv 1\,\, ({\rm mod}\,\, 4)$ every element $c=\alpha^t\in\mathcal{D}$ is a nonsquare since in this case $t$ is always odd. Hence, in this case $-2\bar{c}+c$ is a square in $\gf_{p^n}$ for any $c\in\mathcal{D}$.
		
 \emph{\textbf{Case 2:}} $p^m\equiv 3\,\, ({\rm mod}\,\, 4)$. Then $(-1)^{\frac{p^m+1}{2}}=1$, and $-2\bar{c}+c$ is a square if and only if $c^{\frac{p^n-1}{2}}=1$. In this case $\frac{p^m+1}{6}$ is even, and thus every element $c\in \mathcal{D}$ is a square in $\gf_{p^n}$. So $-2\bar{c}+c$ is also a square in $\gf_{p^n}$ for any $c\in\mathcal{D}$ in this case.	

 Based on the above discussion, one can conclude that for any $c\in\mathcal{D}$, $-2\bar{c}+c$ is a square in $\gf_{p^n}$.
\hfill$\square$

 With the notation introduced above, we stress a simple fact that if $\alpha^{t}$ is a square in $\gf_{p^n}$ with $t$ satisfying $0\leq t<p^n-1$, then $t$ must be even and the square roots of $\alpha^t$ are $\pm \alpha^{t/2}$. In what follows, for a square element $x\in\mathbb{F}_{p^n}^*$, we always
 write it first in the form $x=\alpha^t$ with the restriction $t\in\{0,1,\cdots,p^n-2\}$, and use the convention that  $x^{1/2}=\alpha^{t/2}$. The following proposition gives the value of $\omega_2$.

\begin{proposition}\label{lemmafor-w2}
Let $F(x)=x^{p^m+2}$ be defined as in (\ref{mapping-studied}) and $\mathbb{S}$ be its differential spectrum given in (\ref{dsforoddp}) for $p>3$. When $p^m\equiv 2\,\, ({\rm mod}\,\, 3)$, the component $\omega_2$ in $\mathbb{S}$ is equal to $p^m-1$.
\end{proposition}
{\em Proof:}
As we have shown above, when $p^m\equiv 2\,\, ({\rm mod}\,\, 3)$, $\omega_2$ is equal to the number of $c\in\mathcal{D}$ such that (\ref{det-omega2}) has exactly  two solutions in $\gf_{p^n}$.  In what follows, we will determine $\omega_2$ via studying the equation system (\ref{det-omega2}). For convenience, in the sequel we always denote $p^m$ by $q$.

First we introduce some basic facts. For each $c\in \mathcal{D}$, we have $D=\bar{c}^2-\bar{c}c+c^2=0$, which implies 
$$(-2\bar{c}+c)^2=-3c^2\,\,{\rm and} \,\,(\bar{c}+c)^2=3\bar{c}c.$$
Let $3=\alpha^{t_0}$ for some $t_0\in\{0,1,\cdots,p^n-2\}$ and note that $-1=\alpha^{\frac{q^2-1}{2}}$. We can assume that
\begin{equation}\label{epsilon}
-2\bar{c}+c=\epsilon_c \alpha^{\frac{q^2-1}{4}}\alpha^{\frac{t_0}{2}}c ,
\end{equation}
where $\epsilon_c\in \{-1,1\}$ and its exact value depends on $c$. By the definition of $\mathcal{D}$ in (\ref{defofD}), we have $c=\alpha^{\frac{q+1}{6}}\cdot \alpha^{j(q+1)}$ or $c=\alpha^{\frac{5(q+1)}{6}}\cdot \alpha^{j(q+1)}$, where $j=0,1,\cdots,q-2$. By taking $c=\alpha^{\frac{q+1}{6}}\cdot \alpha^{j(q+1)}$, we have
$$-2\bar{c}+c=\alpha^{j(q+1)}\left(-2\alpha^{\frac{q+1}{6}q}+\alpha^{\frac{q+1}{6}}\right).$$
Together with (\ref{epsilon}) we get  $$-2\alpha^{\frac{q+1}{6}q}+\alpha^{\frac{q+1}{6}}=\epsilon_c \alpha^{\frac{q^2-1}{4}}\alpha^{\frac{t_0}{2}}\alpha^{\frac{q+1}{6}}=\epsilon_c \alpha^{\frac{q^2-1}{4}}3^{\frac{1}{2}}\alpha^{\frac{q+1}{6}},$$
which shows $\epsilon_c$ is only dependent on $\alpha^{\frac{q+1}{6}}$. Similarly, by taking $c=\alpha^{\frac{5(q+1)}{6}}\cdot \alpha^{j(q+1)}$, one can also show that $\epsilon_c$ is only dependent on $\alpha^{5\frac{q+1}{6}}$. Thus, we can conclude that  $\epsilon_c$ in (\ref{epsilon}) is independent of the choice of $j\in\{0,1,\cdots,q-2\}$. 

As for $(\bar{c}+c)^2=3\bar{c}c$, note that $$\bar{c}c=\alpha^{\frac{i(q+1)^2}{6}}\cdot \alpha^{2j(q+1)},$$ where $i=1$ or $5$. Then, we can assume that \begin{equation}\label{tauc}
\bar{c}+c=\tau_c 3^{\frac{1}{2}}\left(\alpha^{\frac{i(q+1)^2}{6}}\right)^{\frac{1}{2}}\alpha^{j(q+1)}
\end{equation} with $\tau_c\in \{\pm1\}$. Similarly, by taking $c=\alpha^{\frac{i(q+1)}{6}}\cdot \alpha^{j(q+1)}$, one can also show that $\tau_c$ is independent of the choice of $j$.

Next it is convenient to distinguish the following two cases.

 \emph{{\textbf{Case 1:}}} $q\equiv 1\,\, ({\rm mod}\,\, 4)$.  Then, $-1$ is a square in $\gf_{q}$. When $q\equiv 2\,\,({\rm mod}\,\, 3)$, we already know that $-3$ is a nonsquare in $\gf_{q}$ by Lemma \ref{quadartic-3}. Thus, $3$ is a nonsquare in $\gf_{q}$ in this case, and we can write it as $3=\alpha^{(q+1)r}$, where $\alpha$ is $r$ is odd. Now take $c=\alpha^{\frac{q+1}{6}+j(q+1)}$, where $j\in \{0, 1, \cdots q-2\}$. Then, we have  $$-2\bar{c}+c=\epsilon_c \alpha^{\frac{q^2-1}{4}}\cdot \alpha^{(\frac{q+1}{2})r}\cdot \alpha^{\frac{q+1}{6}}\cdot \alpha^{j(q+1)},$$
 where $\epsilon_c $ is given in (\ref{epsilon}) and it is independent of the choice of $j$.  By Lemma \ref{lemmasquare}, $\frac{-2\bar{c}+c}{12}$ is always a square in $\gf_{q^2}.$ Thus, from the first equation of (\ref{det-omega2}) we get
 \begin{equation}\label{solutiony}
 y=\pm \left(\frac{-2\bar{c}+c}{12}\right)^{\frac{1}{2}}=\pm \left( \frac{\epsilon_c}{4}\right)^{\frac{1}{2}}\cdot \left(\alpha^{\frac{q^2-1}{4}}\right)^{\frac{1}{2}}\cdot \left(\alpha^{\frac{q+1}{6}(1-3r)}\right)^{\frac{1}{2}}\cdot \left(\alpha^{j(q+1)}\right)^{\frac{1}{2}},
 \end{equation}
   where we use the fact that $3=\alpha^{(q+1)r}$. Now we should check whether this $y$ satisfies the second equation of (\ref{det-omega2}).
   Since $q\equiv 1({\rm mod}\, 4)$, we can assume that $q=4k+1$ for some positive integer $k$, and thus $\frac{q+1}{2}=2k+1$. Next, using $y$ given in (\ref{solutiony}) we compute
\begin{equation}\label{yyq}
\begin{split}
\bar{y}y
&= y^{q+1} \\
&=  \left(\frac{\epsilon_c}{4}\right)^{\frac{q+1}{2}}\cdot \left(\alpha^{\frac{q^2-1}{4}}\right)^{\frac{q+1}{2}}\cdot \left(\alpha^{\frac{q+1}{6}(1-3r)}\right)^{\frac{q+1}{2}} \left(\alpha^{j(q+1)}\right)^{\frac{q+1}{2}}\\
 &= \frac{\epsilon_c}{4}\cdot \left(\alpha^{\frac{q^2-1}{4}}\right)^{2k}\cdot \alpha^{\frac{q^2-1}{4}}\cdot \alpha^{-\frac{q+1}{2}\frac{q+1}{2}r}\cdot \alpha^{\frac{q+1}{6}\cdot \frac{q+1}{2}}\cdot \alpha^{j(q+1)\cdot \frac{q+1}{2}}
\\&=  \frac{\epsilon_c}{4}\cdot (-1)^k\cdot \alpha^{\frac{q^2-1}{4}}\cdot \alpha^{-\frac{q^2-1}{4}r}\cdot \alpha^{-\frac{q+1}{2}r}\cdot \alpha^{\frac{q+1}{6}\cdot \frac{q+1}{2}}\cdot \alpha^{j(q+1)}\cdot (-1)^j
\\&=  \frac{\epsilon_c}{4}\cdot \alpha^{\frac{q^2-1}{4}(1-r)}\cdot \alpha^{-\frac{q+1}{2}r}\cdot \alpha^{\frac{q+1}{6}\cdot \frac{q+1}{2}}\cdot \alpha^{j(q+1)}\cdot (-1)^{j+k}
\\&=  \frac{\epsilon_c}{4}\cdot (-1)^{\frac{1-r}{2}}\cdot \alpha^{-\frac{q+1}{2}r}\cdot \alpha^{\frac{q+1}{6}\cdot \frac{q+1}{2}}\cdot \alpha^{j(q+1)}\cdot (-1)^{j+k}
\\&= \frac{\epsilon_c}{4}\cdot \alpha^{-\frac{q+1}{2}r}\cdot \alpha^{\frac{q+1}{6}\cdot \frac{q+1}{2}}\cdot \alpha^{j(q+1)}\cdot (-1)^{j+k+\frac{1-r}{2}},
\end{split}
\end{equation}
where the sixth equality holds since $r$ is odd.
    Furthermore, we compute $\frac{\bar{c}+c}{12}$.  Since $$3\bar{c}c=\alpha^{(q+1)r}\cdot \alpha^{\frac{q+1}{6}\cdot (q+1)}\cdot \alpha^{j(q+1)^2}=\alpha^{(q+1)r}\cdot \alpha^{\frac{(q+1)^2}{6}}\cdot \alpha^{2j(q+1)},$$ by (\ref{tauc}) we have $$\bar{c}+c=\tau_c \alpha^{\frac{(q+1)r}{2}}\alpha^{\frac{q+1}{6}\cdot \frac{q+1}{2}}\cdot \alpha^{j(q+1)},$$
     where $\tau_c$ is independent of $j$. Then,  we get
\begin{equation}\label{ccq}
 \frac{\bar{c}+c}{12}=\frac{\tau_c}{4}\cdot \alpha^{-\frac{q+1}{2}r}\cdot \alpha^{\frac{q+1}{6}\cdot \frac{q+1}{2}}\cdot \alpha^{j(q+1)}.
\end{equation}
Comparing (\ref{yyq}) with (\ref{ccq}), we conclude that for each $c=\alpha^{\frac{q+1}{6}+j(q+1)}$, $j\in \{0, 1, \cdots q-2\}$, the values of $y$ given by (\ref{solutiony}) are solutions of (\ref{det-omega2}) if and only if
\begin{equation}\label{condifortwo}(-1)^{j+k+\frac{1-r}{2}}=\epsilon_c\tau_c,\end{equation} where $\epsilon_c,\tau_c\in \{-1,1\}$. Note that since $k$, $r$ are fixed positive integers and $\epsilon_c$, $\tau_c$ are independent of the choice of $j$. Thus, when $j$ runs through $\{0,1,\cdots,q-2\}$, half of these $j$'s satisfy (\ref{condifortwo}). This shows that for $c=\alpha^{\frac{q+1}{6}+j(q+1)}$ with $j\in \{0, 1, \cdots q-2\}$, there are half of these $c$'s such that (\ref{det-omega2}) has exactly  two solutions in $\gf_{p^n}$. Similarly, in this case for $c=\alpha^{\frac{5(q+1)}{6}+j(q+1)}$ with $j\in \{0, 1, \cdots q-2\}$, we can obtain the same result.

  \emph{{\textbf{Case 2:}}} $q\equiv -1\,\, ({\rm mod}\,\, 4)$. Then, $-1$ is a nonsquare in $\mathbb{F}_q$. Next, without loss of generality we take $c=\alpha^{\frac{5(q+1)}{6}+j(q+1)}$ with $j\in \{0, 1, \cdots q-2\}$. For $c=\alpha^{\frac{(q+1)}{6}+j(q+1)}$ with $j\in \{0, 1, \cdots q-2\}$, one can derive the same result.
  First assume that $3=\alpha^{(q+1)s_0}$, where $s_0$ is even since $3$ is a square in $\mathbb{F}_q$ in this case. Then, we have
  $-2\bar{c}+c=\epsilon_c \alpha^{\frac{q+1}{2}(\frac{q-1}{2}+s_0)}\alpha^{\frac{5(q+1)}{6}+j(q+1)}$, where $\epsilon_c$ is given in (\ref{epsilon}). Denote $\frac{q-1}{2}+s_0\,\,{\rm mod}\,\,2(q-1)$ by $s$. We can
  conclude that $s$ is odd since $\frac{q-1}{2}$ is odd.  Note that $\alpha^{\frac{q+1}{2}s}=-3$. From the first equation of (\ref{det-omega2}) we obtain
 \begin{equation}\label{solutiony0}
 y=\pm \left(\frac{-2\bar{c}+c}{12}\right)^{\frac{1}{2}}=\pm \left(-\frac{\epsilon_c}{4}\right)^{\frac{1}{2}}\cdot\left(\alpha^{-\frac{q+1}{2}s}\right)^{\frac{1}{2}}\cdot \left(\alpha^{\frac{5(q+1)}{6}}\right)^{\frac{1}{2}}\cdot\left(\alpha^{j(q+1)}\right)^{\frac{1}{2}}.
 \end{equation}
 Then, using $y$ in (\ref{solutiony0}) we compute
\begin{equation}\label{yyq0}
\begin{split}
\bar{y}y
&= y^{q+1} \\
&=  \left(-\frac{\epsilon_c}{4}\right)^{\frac{q+1}{2}}\cdot \left(\alpha^{-\frac{q+1}{2}s}\right)^{\frac{q+1}{2}}\cdot \left(\alpha^{\frac{5(q+1)}{6}}\right)^{\frac{q+1}{2}} \left(\alpha^{j(q+1)}\right)^{\frac{q+1}{2}}\\
 &= \frac{1}{4}\cdot \alpha^{-\frac{q^2-1}{4}s}\cdot \alpha^{-\frac{q+1}{2}s}\cdot \alpha^{\frac{5(q+1)}{6}\cdot \frac{q+1}{2}}\cdot \alpha^{j(q+1)}(-1)^j
\\&=  \frac{1}{4}\cdot (-1)^{\lfloor \frac{-s}{2}\rfloor}\cdot \alpha^{\frac{q^2-1}{4}}\cdot \alpha^{-\frac{q+1}{2}s}\cdot \alpha^{\frac{5(q+1)}{6}\cdot \frac{q+1}{2}}\cdot \alpha^{j(q+1)}\cdot (-1)^j,
\end{split}
\end{equation}
where $\lfloor \frac{-s}{2}\rfloor$ denotes the greatest integer less than  or equal to $\frac{-s}{2}$.
   On the other hand, by (\ref{tauc}) we have
    \begin{equation}\label{cc12}
 \frac{\bar{c}+c}{12}=(-\frac{\tau_c}{4})\alpha^{\frac{q^2-1}{4}}\cdot\alpha^{-\frac{(q+1)}{2}s}\cdot\alpha^{\frac{5(q+1)}{6}\frac{q+1}{2}}\cdot\alpha^{j(q+1)}.
 \end{equation}
 Comparing (\ref{yyq0}) with  (\ref{cc12}), the values of $y$ in  (\ref{solutiony0}) are solutions of (\ref{det-omega2}) if and only if
  $$(-1)^{j+\lfloor \frac{-s}{2}\rfloor}=-\tau_c.$$
  For given  $q$, $\lfloor \frac{-s}{2}\rfloor$ is a fixed integer and $\tau_c\in\{\pm1\}$ is only dependent on $\alpha^{\frac{5(q+1)}{6}}$. Thus,  one can conclude that when $j$ runs through $\{0,1,\cdots,q-2\}$, the number of  $j$ satisfying the above equation is equal to $\frac{q-1}{2}$.
  For $c=\alpha^{\frac{(q+1)}{6}+j(q+1)}$ with $j\in \{0, 1, \cdots q-2\}$, we can deduce the same result. Thus, in this case for $c\in \mathcal{D}$, there are $2\cdot\frac{q-1}{2}$ $c$'s such that (\ref{det-omega2}) has exactly two solutions in $\gf_{p^n}$.

  Based on the above discussions, we conclude that for given $q=p^m$ satisfying $p^m\equiv 2\,\, ({\rm mod}\,\, 3)$, half of the elements in $\mathcal{D}$ make (\ref{det-omega2}) have exactly two solutions in $\gf_{p^n}$. Thus, we obtain $\omega_2=q-1$.
 \hfill$\square$

With the above preparations, we determine the differential spectrum of $F(x)=x^{p^m+2}$ for $p>3$. The main results are given in the following theorem.

\begin{theorem}\label{diff-spec}
Let $F(x)$ be the power mapping over $\gf_{p^n}$ defined as in (\ref{mapping-studied}) with $p>3$. When $p^m\equiv 1\,\, ({\rm mod}\,\, 3)$, $F(x)$
is a differentially $2$-uniform mapping and its differential spectrum is given by
$$
\mathbb{S}=\left[\omega_0=\frac{p^n-1}{2},\,\,\omega_1=1,\,\,\omega_2=\frac{p^n-1}{2}\right].
$$
When $p^m\equiv 2\,\, ({\rm mod}\,\, 3)$, $F(x)$ is a differentially $4$-uniform mapping and the differential spectrum is given by
$$\mathbb{S}=\left[\omega_0=\frac{(3p^m+1)(p^m-1)}{4}, \,\,\omega_1=1,\,\, \omega_2=p^m-1,\,\, \omega_3=0,\,\, \omega_4=\frac{(p^m-1)^2}{4}\right].$$
\end{theorem}
{\em Proof:}  When $p>3$, let $\mathbb{S}$ be the differential spectrum of $F(x)$ defined as in  (\ref{dsforoddp}). By the discussions presented there, we have shown that $\omega_1=1$ and $\omega_3=0$. Moreover, when $p^m\equiv 1\,\, ({\rm mod}\,\, 3)$,  we also have $\omega_4=0$.

{\em \textbf{Case 1:}} $p^m\equiv 1\,\, ({\rm mod}\,\, 3)$. Then, the power mapping $F(x)$ over $\mathbb{F}_{p^n}$ is differentially $2$-uniform. Since $\omega_1=1$, according to the  identities in (\ref{prop}), we have
\begin{equation*}\arraycolsep=1.2pt\def\arraystretch{1.7}
\left \{\begin{array}{lll}
\omega_0+\omega_1+\omega_2=p^n,\\
\omega_1+2\omega_2=p^n,\\
\omega_1=1.
\end{array}\right.\end{equation*}
Solving this equation system, the differential spectrum of $F(x)$ is obtained.
	
{\em \textbf{Case 2:}} $p^m\equiv 2\,\, ({\rm mod}\,\, 3)$. In this case we have known that $\omega_1=1$ and $\omega_3=0$. Moreover, by Proposition \ref{lemmafor-w2} we also have $\omega_2=p^m-1$.  Then, still utilizing the identities in (\ref{prop}), we can obtain a equation system in unknowns $\omega_0$ and $\omega_4$. Solving the obtained equation system, we get the desired result.
\hfill$\square$

Utilizing the differential spectrum of $F(x)$ in Theorems \ref{dsforp=2}, \ref{dsforp=3} and \ref{diff-spec}, the ambiguity and deficiency of $F(x)$ can be easily computed. Here we only give the corresponding result
for the case $p>3$.

\begin{corollary}
Let $F(x)=x^{p^m+2}$ be the power mapping over $\gf_{p^n}$ with $p>3$. The ambiguity and deficiency of $F(x)$ are denoted by $A(F)$ and $D(F)$, respectively. When $p^m\equiv 1\,\, ({mod}\,\, 3)$, we have $A(F)=\frac{(p^n-1)^2}{2}$ and $D(F)=\frac{(p^n-1)^2}{2}$. When $p^m\equiv 2\,\, ({mod}\,\, 3)$, $A(F)=\frac{\left(p^n-1\right)\left(p^m-1\right)\left(3p^m-1\right)}{2}$ and $D(F)=\frac{\left(p^n-1\right)\left(p^m-1\right)\left(3p^m+1\right)}{4}$.
\end{corollary}
{\em Proof:} The results are obtained by direct computations. When $p^m\equiv 1\,\, ({\rm mod}\,\, 3)$, $F(x)=x^{p^m+2}$ is  differentially $2$-uniform. Then, by the differential spectrum of $F(x)$ in Theorem \ref{diff-spec}, we obtain the ambiguity $A(F)$
 and the deficiency $D(F)$ of $F(x)$ as follows:

   $$A(F)=(p^n-1)\omega_2\binom{2}{2}=(p^n-1)\omega_2=\frac{(p^n-1)^2}{2},$$ and
$$D(F)=(p^n-1)\omega_0=\frac{(p^n-1)^2}{2}.$$
When $p^m\equiv 2\,\, ({\rm mod}\,\, 3)$, we have
$$A(F)=(p^n-1)\sum\limits_{i=2}^{4} \omega_i\binom{i}{2}=(p^n-1)(\omega_2+6\omega_4)=\frac{\left(p^n-1\right)\left(p^m-1\right)\left(3p^m-1\right)}{2},$$ and
$$D(F)=(p^n-1)\omega_0=\frac{\left(p^n-1\right)\left(p^m-1\right)\left(3p^m+1\right)}{4}.$$	
\hfill$\square$

\section{Some results about the differential properties of $x^{p^m+2}$ when $n=2m-1$}\label{result-in-odd}
We have studied the differential properties of the power mapping $F(x)=x^{p^m+2}$ over $\gf_{p^n}$ when $n=2m$. A natural question one would ask is whether the power mapping $F(x)=x^{p^m+2}$ over $\gf_{p^n}$ has similar differential properties when $n=2m-1$. The aim of this section is to investigate this question.

Now we assume that  $n=2m-1$ and $p$ is a prime. When $p=2$, by Remark \ref{rem1}, the power mapping $F(x)=x^{p^m+2}$ over $\gf_{p^n}$ is differentially $2$-uniform (in this case $F(x)$ is called an almost perfect nonlinear function ) and its differential spectrum is $$[\omega_0=2^{n-1}, \omega_1=0, \omega_2=2^{n-1}].$$

  Next we assume that  $p$ is an odd prime. Setting $c=4b-1$, $y=x+\frac{1}{2}$ and $\bar{y}=y^{p^m}$, the derivative equation $\mathbb{D}_1F(x)=b$ of $F(x)=x^{p^m+2}$ over $\gf_{p^n}$
  can  be transformed into
  \begin{equation}\label{diff-equa3-nodd}
8y\bar{y}+4y^2-c=0.
\end{equation}
 When $c=0$, (\ref{diff-equa3-nodd})  implies that $y^{p}=\frac{1}{4}y$, which has exactly one solution $y=0$ or  $p$ solutions in $\mathbb{F}_{p^n}$. Note that by that fact $y^{p}=\frac{1}{4}y$, we can derive that $\bar{y}=\frac{1}{4^m}y$. Substituting this into  (\ref{diff-equa3-nodd}), we can conclude that (\ref{diff-equa3-nodd}) has  $p$ solutions in $\mathbb{F}_{p^n}$ if and only if $\frac{1}{4}$ is a $(p-1)$th power in $\mathbb{F}_{p^n}$ and $1+2^n$ is equal to zero in $\mathbb{F}_{p}$. Otherwise,  (\ref{diff-equa3-nodd}) has exactly one solution, that is $y=0$. If $c\neq 0$, then $y\neq 0$ and  from (\ref{diff-equa3-nodd}) we further have
\begin{equation}\label{ybar-nodd}
\bar{y}=\frac{-y^2+\frac{c}{4}}{2y}.
\end{equation}
Raising (\ref {diff-equa3-nodd}) to the  power $p^m$, we obtain
\begin{equation}\label{diff-equa4-nodd}
8\bar{y}y^p+4\bar{y}^2-\bar{c}=0.
\end{equation}
Substituting $\bar{y}$ in (\ref{ybar-nodd}) into (\ref {diff-equa4-nodd}), we have
\begin{equation}\label{diff-equa5-nodd}
64y^{p+3}-16cy^{p+1}-16y^4+(8c+16\bar{c})y^2-c^2=0.
\end{equation}
The degree of equation (\ref{diff-equa5-nodd}) is $p+3$, thus it has at most $p+3$ solutions in $\mathbb{F}_{p^n}$ and so does (\ref{diff-equa3-nodd}). Thus, the differential uniformity  $\delta(F)$ of $F(x)=x^{p^m+2}$ is less than or equal to $p+3$.  Moreover, note that when $c\neq 0$, the solutions of (\ref{diff-equa3-nodd}) (resp. (\ref{diff-equa5-nodd})) come in pairs. Therefore, when $c\neq 0$, the possible numbers of solutions of (\ref{diff-equa3-nodd}) (resp. (\ref{diff-equa5-nodd}))
are $0, 2, 4,\cdots,p+3$, which are exactly the even integers between $0$ and $p+3$.
We summarize these results in the following theorem.

\begin{theorem}\label{result for odd case} Let $p$ be an odd prime and $n=2m-1$. Denote the differential spectrum of the power mapping $F(x)=x^{p^m+2}$ over $\mathbb{F}_{p^n}$ by $[\omega_0,\omega_1,\cdots,\omega_{p+3}]$. Then, we have the following results: 

\noindent (\textrm{i}) the differential uniformity of  $F(x)=x^{p^m+2}$ is less than or equal to $p+3$;  

\noindent (\textrm{ii}) the component $\omega_i=0$ if $i$ is odd and $i\notin \{1,p\}$, where $0<i<p+3$;
 
 \noindent (\textrm{iii}) if $\frac{1}{4}$ is a $(p-1)$th power in $\mathbb{F}_{p^n}$ and $1+2^n=0$ in $\mathbb{F}_{p}$, then $\omega_p=1$ and $\omega_1=0$. Otherwise, $\omega_p=0$ and $\omega_1=1$.
\end{theorem}

According to Theorem \ref{result for odd case}, in order to compute the differential spectrum of $F(x)=x^{p^m+2}$ in the case $n=2m-1$, we need to determine the values of $\omega_i$ with $i$ being even and $0\leq i\leq p+3$. It seems difficult to achieve this goal since the involved  equations (\ref{diff-equa3-nodd}) and (\ref{diff-equa5-nodd}) have higher degree as the value of $p$ increases, and we cannot find an efficient method to deal with them. Next we provide a numerical example.
\begin{example}Let $p=5$ and $m=4$. By \textbf{Magma}, the differential spectrum of $F(x)=x^{p^m+2}$ over $\mathbb{F}_{5^7}$ can be directly computed:
$$[\omega_0=47630,\omega_1=1,\omega_2=22710,\omega_3=0,\omega_4=7392,\omega_5=0,\omega_6=0,\omega_7=0,\omega_8=392],$$
which coincides with the statements in Theorem \ref{result for odd case}.
\end{example}

For the case $n=2m-1$, there are two many unknown components in the differential spectrum of $F(x)=x^{p^m+2}$ as the prime $p$ increases. Thus, it is challenging to determine the differential spectrum completely in general. When $p=3$, note that $3^m+2$ and   $2\cdot 3^{m-1}+1$ are in the same cyclotomic coset of modulo $3^n-1$. Then, the power mappings $x^{2\cdot 3^{m-1}+1}$ and $F(x)=x^{3^m+2}$ over $\mathbb{F}_{3^n}$ share the same differential spectrum.  Dobbertin et al. presented a delicate and complicated method for computing the differential spectrum of $x^{2\cdot 3^{m-1}+1}$  in \cite{DHK2011IT}, where $2\cdot 3^{m-1}+1$ was called the Welch exponent. According to the results of \cite{DHK2011IT}, we can obtain the differential spectrum of $F(x)=x^{3^m+2}$ as follows.

\begin{theorem}(\cite{DHK2011IT})
Let $F(x)=x^{3^m+2}$ be the power mapping over $\mathbb{F}_{3^n}$ with $n=2m-1$. Then $F(x)$ is a differentially $4$-uniform mapping and its differential spectrum is given by
$$\mathbb{S}=[\omega_0=\frac{5\cdot 3^n+1}{8},\,\,\omega_1=0,\,\,\omega_2=\frac{3^n-3}{4},\,\,\omega_3=1,\,\,\omega_4=\frac{3^n-3}{8}].$$
\end{theorem}

We are not sure whether the method in \cite{DHK2011IT} for $p=3$ can be generalized to other odd primes, but it is really an interesting and challenging problem to determine the differential spectrum of $F(x)=x^{p^m+2}$ in Theorem \ref{result for odd case}. The readers are invited to contribute to this problem.

\section{Conclusion}\label{con-remarks}
In this paper, for any prime $p$, we conduct a comprehensive investigation on the differential spectrum of the power mapping $F(x)=x^{p^m+2}$ over $\gf_{p^n}$. For the case $n=2m$, we propose some new techniques to deal with the derivative equation of $F(x)$, and thus the differential spectrum is completely computed. Our results can be regarded as an improvement of that in \cite{HRS1999IT}. For the case $n=2m-1$, we derive partial results about the differential spectrum of $F(x)=x^{p^m+2}$. Since the involved equations in this case have higher degree, it seems difficult to calculate the differential spectrum. We leave this as an open problem. The exponent studied in this paper is very interesting: it is the Gold exponent when $p=2$, the Niho exponent when $p=3$ and $n=2m$, and the ternary Welch exponent when $p=3$ and $n=2m-1$.

\section*{Acknowledgment}
Y. Xia and Y. Man were supported in part by the National Natural
Science Foundation of China under Grants 62171479, 61771021, and in part by the Fundamental Research Funds for the Central Universities, South-Central University for Nationalities under Grant CZT20023. C. Li and T. Helleseth were supported by the Research Council of Norway under Grants 247742 and 311646.

\end{document}